\documentclass[noshowpacs,twocolumn,amssymb,preprintnumbers,prc,floatfix,amsmath]{revtex4-1}
\usepackage{rotating,epsfig,dcolumn}
\usepackage{graphicx,bm}
\usepackage{bm,color}
\usepackage{epstopdf}
\newcommand{\A}[1]{$^{#1}$}
\newcommand{\vlk}{V_{\text{low-}k}}

\newcommand{\bL}{\begin{Large}}
\newcommand{\eL}{\end{Large}}
\newcommand{\be}{\begin{equation}}  
\newcommand{\ee}{\end{equation}}
\newcommand{\ba}{\begin{eqnarray*}}
\newcommand{\ea}{\end{eqnarray*}}

\newcommand{\hw}{$\hbar \omega \,$}
\newcommand{\hbw}{\hbar \omega}

\newcommand{\ie}{{\it i.e.,\ }}

\newcommand{\MDE}{\text{MDE}}

\begin{document}
\title{Neutron Skins and Halo Orbits in the $\bm{sd}$ and $\bm{pf}$ shells}
\author{J.~Bonnard$^1$, S.~M.~Lenzi$^{1,2}$ and A.~P.~Zuker$^{2,3}$}
\affiliation{$^1$ Istituto Nazionale di Fisica Nucleare, Sezione di
  Padova, 35131 Padova, Italy\\ $^2$ Dipartimento di Fisica e
  Astronomia, Universit\`a degli Studi di Padova, I-35131 Padova,
  Italy\\$^3$ Universit\'e de Strasbourg, IPHC, CNRS, UMR7178, 23 rue
  du Loess 67037 Strasbourg, France}
\begin{abstract}
The strong dependence of Coulomb energies on nuclear radii makes it
possible to extract the latter from calculations of the former. The
resulting estimates of neutron skins indicate that two mechanisms are
involved. The first one---isovector monopole polarizability---amounts
to noting that when a particle is added to a system it drives the
radii of neutrons and protons in different directions, tending to
equalize the radii of both fluids independently of the neutron excess.
This mechanism is well understood and the Duflo-Zuker (small) neutron
skin values derived 14 years ago are consistent with recent measures
and estimates. The alternative mechanism involves halo orbits whose
huge sizes tend to make the neutron skins larger and have a subtle
influence on the radial behavior of $sd$ and $pf$ shell nuclei. In
particular, they account for the sudden rise in the isotope shifts of
nuclei beyond $N=28$ and the near constancy of radii in the $A=40-56$
region. This mechanism, detected here for the first time, is not well
understood and may well go beyond Efimov physics usually associated to
halo orbits. 
\end{abstract}
\date{\today}
\maketitle 

Mirror nuclei in which proton and neutron numbers $N, Z$ are interchanged
have different energetics due to the isospin breaking interactions
(IBI) dominated by the Coulomb force. It affects both the spectra (MED
for Mirror Energy Differences) and the ground states (MDE for Mirror
Displacement Energies). A prime example of the MED is found in $^{13}$Ni
where the $1s_{1/2}$ proton orbit is depressed by about 750 keV with
respect of its neutron analogue in $^{13}$C. This behavior is referred
to as Thomas-Ehrman shift (TES) because it was first studied via
R-matrix theories by J. Ehrman~\cite{ehrman} and
R. G. Thomas~\cite{thomas} who also considered the pair $^{17}$F-$^{17}$O.

The $s$ orbits are the essential ingredients of halo
physics~\cite{halo} and have a decisive influence in the spectroscopy
of nuclei at $A=16\pm 1$~\cite{OHNH99,GGZ15}, which will be shown to
extend to higher masses, including the $pf$ shell because of the halo
nature of the $p$ orbits.

The TES can be viewed as an overbinding of orbits---with respect to
naive expectations---because their large radii reduce the Coulomb
repulsion. For the binding energies, the naive assumption is that a
closed shell core is unperturbed by the addition of a particle. The
MDE would then be due to the core Coulomb field acting on the extra
proton. The result is often a severe underestimate, as in $A=41$: the
Nolen-Schiffer anomaly (NSA)~\cite{NS69} illustrated in
Table~\ref{tab:isos}.
\begin{table}[h]
\caption{\label{tab:isos}Displacement energies between the ground
  state of $T=1/2$ mirror nuclei of mass $A$ defined as
  $\MDE=E_{J}(Z>N)-E_J(Z<N)$. Experimental, full IBI, Coulomb (C) and
  schematic Coulomb (Eq.~(\ref{mde}), sC) contributions are given in
  MeV. No core 0\hw calculation with $\vlk$ form~\cite{vlowk} of the
  N3LO~\cite{n3lo} potential with cutoff $\lambda=2.0$ fm$^{-1}$.}
\begin{tabular*}{\linewidth}{@{\extracolsep{\fill}}ccccccc}
\hline
\hline
$A$& $\hbw$& $J^\pi$& $\MDE{}_{\text{exp}}$& $\MDE{}_{\text{IBI}}$& $\MDE{}_{\text{C}}$& $\MDE{}_{\text{sC}}$\\ 
\hline
15   & 14.67 & $1/2^-$ & 3.537 & 3.574 & 3.474 & 3.624 \\ 
17   & 13.38 & $5/2^+$ & 3.543 & 3.388 & 3.377 & 3.514 \\
39   & 10.89 & $3/2^+$ & 7.307 & 7.120 & 6.970 & 7.212 \\
41   & 10.61 & $7/2^-$ & 7.278 & 6.683 & 6.679 & 6.675 \\
\hline
\hline
\end{tabular*}
\end{table}
While the TES is due to a proton radius larger than expected for the
extra particle, the NSA may be thought to demand the opposite \ie a
reduction of the radius of the added particle but this is ruled out
experimentally~\cite{Pla88}.

Though Shlomo had noticed that equalizing the total neutron and proton
radii would eliminate the anomaly~\cite{shlomo73} it took some time
before this basically sound idea gained traction: Hartree-Fock (HF)
calculations routinely predicted proton radii in agreement with
experiment and substantially larger neutron radii~\cite{shlomo77},
though experimental evidence did not support the
latter~\cite{shlomoF,varmaz}. Then it was shown that good proton radii
were compatible with a variety of neutron radii~\cite{gomezm,brownEOS}
and calculations appeared in which the NSA was almost
absent~\cite{Agr01}. The NSA does not seem to have attracted much
attention lately but neutron radii are a very hot subject whose
connection with displacement energies---hitherto somewhat neglected---is
worth examining. It follows by noting that isospin conservation implies
that the proton rms radius $\rho_{\pi>}=\sqrt{\langle
  r_{\pi>}^2\rangle}$ of a nucleus with $Z>N$ equals the neutron rms
radius of its mirror, $\rho_{\nu<}$ with $Z<N$. Assuming a schematic
Coulomb contribution of the form $C_{Zx}=0.67Z(Z-1)/\rho_x$ we have
(disregarding other IBI terms)
\begin{gather} \label{mde}
\MDE=C_{Z+1\pi>}-C_{Z\pi<}=C_{Z+1\nu<}-C_{Z\pi<}
\end{gather}
 Therefore, if we know, say, $\MDE(^{17}$F-$^{17}$O) and $\rho_{\pi<}$, 
 the proton radius of $^{17}$O, we also know its neutron radius
 $\rho_{\nu<}$. This simple idea lead to a general estimate of the
 neutron skins by Duflo and Zuker (DZ)~\cite{DZIII}. They started by
fitting the proton mean square radii to experiment
through ($t=N-Z$) 
 \begin{gather}\label{radii}
 \sqrt{\langle r_{\pi}^2\rangle}=\rho_{\pi}=
A^{1/3}\left(\rho_0-\frac{\zeta}{2}\frac{t}{A^{4/3}}-
\frac{\upsilon}{2}(\frac{t}{A})^2\right)e^{(g/A)}\\
+\lambda[z(D_{\pi}-z)/D_{\pi}^2\times n(D_{\nu}-n)/D_{\nu}^2]A^{-1/3}\label{lambda}     
\end{gather}
where $n,\, z$ are the number of active particles between the EI magic
numbers~\cite[Sec. IC]{rmp} at $N,\, Z= $6, 14, 28, 50\ldots; $D_x=8$, 14, 22\ldots are the
corresponding degeneracies.  By fitting known radii for $A\le 60$ one
obtains rms deviations of about 42 mf for a 4 parameter fit with
$\lambda=0$ reduced to about 18 mf when varying $\lambda$. (Including
all known radii the rms deviation goes down, with little change in the
parameters). In principle the neutron skin (in fm)
\begin{equation}\label{zeta}
\Delta r_{\nu\pi}=\rho_\nu-\rho_\pi=\frac{\zeta t}{A}e^{g/A},
\end{equation}
could be expected to come out of the fit. However, fixing $\zeta$
to values between 0.4 and 1.2 did not alter the quality of the fit. A
useful reminder that the neutron radii are independent of the proton
ones.  Hence, the authors resorted to Eq.~(\ref{mde}) using a form of
the Coulomb potential close to the exact one for oscillator orbits.
We adopt the set $g=0.985,\, \rho_0=0.944,\,
\lambda=5.562,\,\upsilon=0.368,\, \zeta=0.8$,\, rmsd=0.0176. All units
in fm except $g$. With these values of $g$ and $\zeta$ Eq.~(\ref{zeta})
yields the estimates of Table~\ref{tab:estim} where they are seen to
agree with numbers of diverse origin: a recent measure~\cite{Pbskin},
estimates based on comparison with electric dipole polarizability
$\alpha_D$~\cite{rocam} and an ``{\em ab initio}''
calculation~\cite{hagen}.
\begin{table}[h]
\caption{\label{tab:estim}Comparing $\Delta r_{\nu\pi}$  from Eq.~(\ref{zeta})
  with estimates (ests)~\cite{hagen,rocam} and
  measure (exp)~\cite{Pbskin} (fm).}
\begin{tabular*}{\linewidth}{@{\extracolsep{\fill}}cccccc}
\hline
\hline
$ $& \A{_{48}}Ca & \A{_{68}}Ni& \A{_{120}}Sn &\A{_{208}}Pb & \A{_{128}}Pb\\ 
\hline
Eq.~(\ref{zeta})&0.14     &0.14   &0.13   &0.17   &0.17\\
ests-exp       &0.135(15)&0.17(2)&0.14(2)&0.16(3)&0.15(3)\\
ref.           &\cite{hagen}     &\cite{rocam}   &\cite{rocam}   &\cite{rocam}   &\cite{Pbskin}\\
\hline
\hline
\end{tabular*}
\end{table}  
It should be noted (stressed) that the results of Eq.~(\ref{zeta})
also square nicely with those obtained from two other sources analyzed
in \cite{warda}: they are very close to the Gogny D1S force~\cite{D1S}
and not far from those of Sly4~\cite{SLy4}---which gives slightly
bigger skins. It appears that a general mechanism, that we sketch
next, is at play. Think of a model space in which an extra particle
(dot in Fig~\ref{fig:pol} taken to be a neutron) associated to number
$n$ and isospin $t$ polarizes the system by inducing particle-hole
jumps from the closed core of particles $h$ to the open shells of
particles $p$ 2\hw above. While $H_0$ represents isoscalar monopole
polarizability, responsible for an overall increase in radius, its
isovector counterpart, $H_1$ takes care of a differential
contraction-dilation of the fluids. The model could be termed the
``degree zero'' of the mean field~\cite{czech}:
\begin{figure}[b]
\begin{center}
\includegraphics[width=0.3\textwidth]{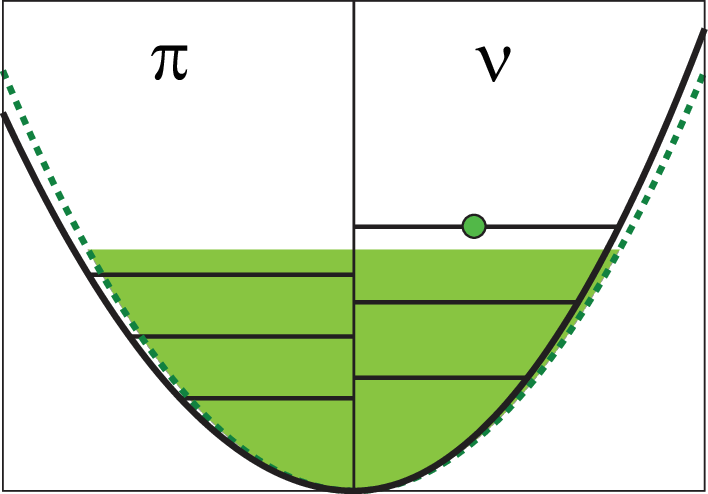}
\end{center}
\caption{\label{fig:pol}(color online) Illustrating the solution of
  Eq.~(\ref{h1}). Explained in text.}
\end{figure}
\begin{gather}
H_0=\varepsilon S_0+v_0n(S_++S_-),\quad \varepsilon=\frac{1}{2}(\varepsilon_p -\varepsilon_h)\label{h0},\\  
S_0=\hat{n_p}-\hat{n_h},\quad S_+=a_p^+a_h+b_p^+b_h+hc,\nonumber\\
H_1=\varepsilon S_0+v_1(t_-U_+ +t_+U_- +\frac{1}{2}t_0U_0),\label{h1}\\
U_0=a_p^+a_h-b_p^+b_h+a_h^+a_p-b_h^+b_p\nonumber,\\
U_+=a_p^+b_h+a^+_hb_p , \qquad U_-=b^+_pa_h+b^+_ha_p\nonumber .
\end{gather}
A unitary (HF) transformation solves exactly $H_0$ but only
approximately $H_1$ because the term in $t_-U_++t_+U_-$ demands a more
refined treatment, ignored here. The results can be visualized in
Fig.~\ref{fig:pol}. The shaded area corresponds to the unperturbed
Hamitonian bounded by a parabola, while the heavy lines represent
parabolic segments with $\hbw_{\nu}>\hbw_{\pi}$, the situation in
which the NSA disappears as the radii tend to equalize \ie reduce the
neutron skin with respect to the $\hbw_{\nu}=\hbw_{\pi}$ value.  The
sign of $v_1$ determines whether radii equalize or move apart. Within
this elementary mean field approach all orbits behave in the same
way. A more refined approach would allow different polarizabilities
for different orbits. Moreover, the model ignores threshold effects
\ie coupling to the continuum that could play an important role.

\begin{figure}[tbh]
\begin{center}
  \includegraphics[width=0.48\textwidth]{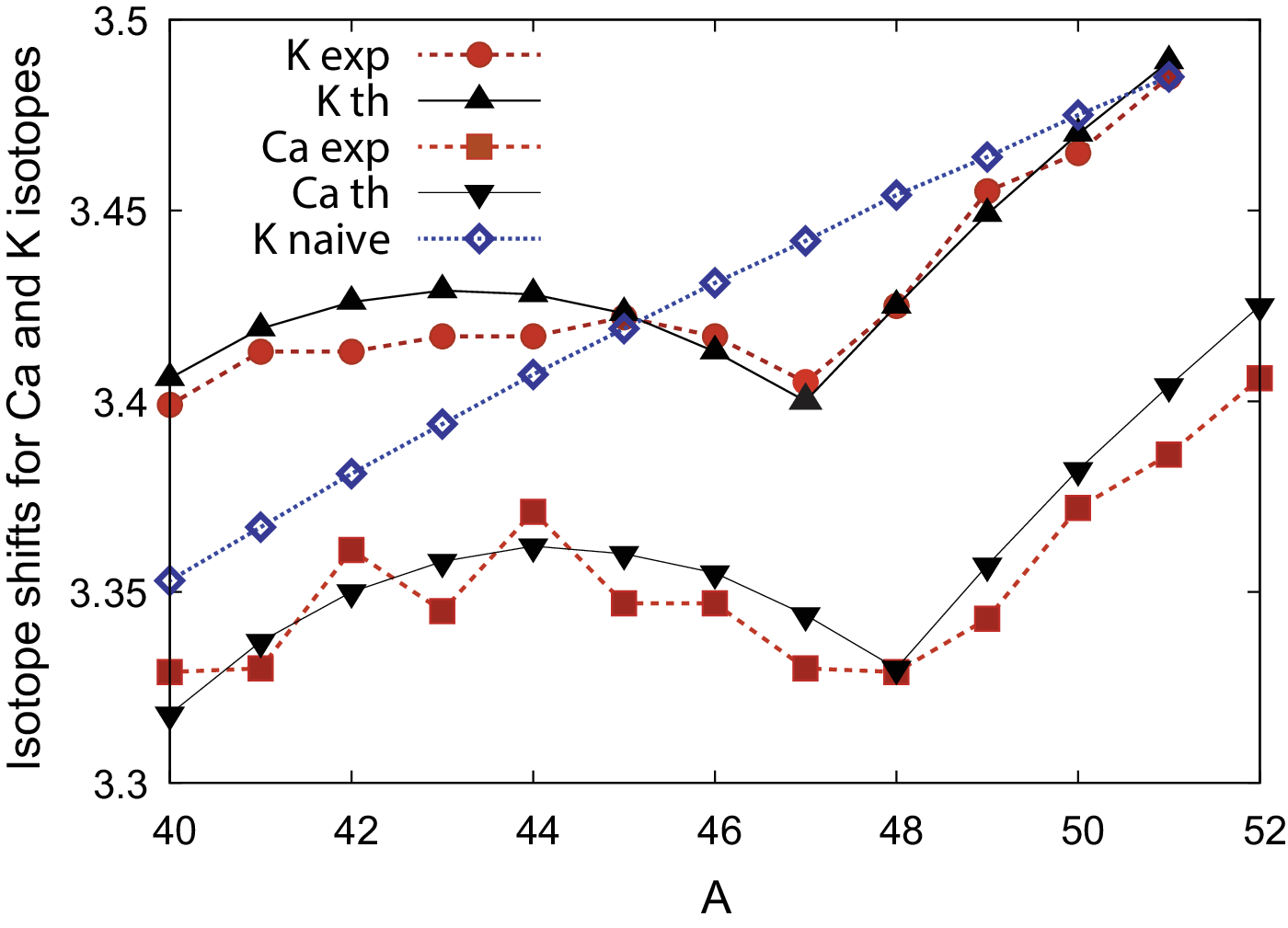}
\end{center}
\caption{\label{fig:CaK} (color online) Radii $\rho_{\pi}$ in fm from
  isotope shifts in the K and Ca isotopes~\cite{AM} incorporating
  recent measures~\cite{Kradii,Caradii}(label exp) compared with
  estimates from Eqs.(2,3) (parameters as in Fig. 2, label naive is
  for $\lambda=0$). The correlated numbers (label th) have been
  shifted down by 30 mf, to restore translation invariance and allow
  for experimental uncertainties in the extraction of radii
  $\rho_{\pi}$ from $\delta \langle r_{\pi}^2\rangle$ isotope
  shifts. For clarity the K and Ca values are shifted by $\pm 55$ mf
  respectively}
\end{figure}

Nonetheless the model has the advantage of suggesting the
computational strategy that generalizes the DZ approach. We shall work
in 0\hw no-core spaces with $\vlk$~\cite{vlowk} precision potentials:
AV18~\cite{AV18}, CDB~\cite{CDB}, and N3LO~\cite{n3lo} which produce
almost indistinguishable results according to our checks. They
incorporate effects not treated in DZ (such as electromagnetic
spin-orbit coupling) are fully IBI and will make it possible to do
configuration mixing.  Saturation is treated in the standard shell
model way by fixing \hw at a value consistent with the observed
radius. It is here that Fig.~\ref{fig:pol} comes in: For each nucleus,
calculations are done for a different \hw for neutrons and protons:
$\hbw_{\pi}$ is known through Eqs.~(\ref{radii},\ref{lambda},\ref{hbw})
for $N>Z$ (and hence $\hbw_{\nu}$ for $N<Z$). Then $\hbw_{\nu}$
for $N>Z$ and $\hbw_{\pi}$ for $N<Z$ follow from $\zeta$ treated as a
free parameter to reproduce the experimental MDE or MED.  To relate
$\hbw_{\pi}$ to the radii we adapt from \cite[Eq.(2.157)]{bm}
Eq.~(\ref{hbw}), where the sum runs over occupied proton orbits in
oscillator shells of principal quantum number $p$, and a similar
expression for neutrons, leading asymptotically to
Eq.~(\ref{hbwasymp}).
\begin{gather}
\hbw_{\pi}=\frac{41.47}{\langle r_{\pi}^2 \rangle}\sum_i z_i(p_i+3/2)/Z  \label{hbw},\\
\frac{\hbar\omega_{\pi}}{(2Z)^{1/3}}=\frac{35.59}{\langle r^2_\pi\rangle};~~~~
\frac{\hbar\omega_{\nu}}{(2N)^{1/3}}=\frac{35.59}{\langle r^2_\nu\rangle}.\label{hbwasymp}
\end{gather}
The form of \hw as a function of $A$ is obtained through a term by
term (nucleus by nucleus) evaluation of Eq~.(\ref{hbw}).
\begin{figure}[t]
\begin{center}
 \includegraphics[width=0.48\textwidth]{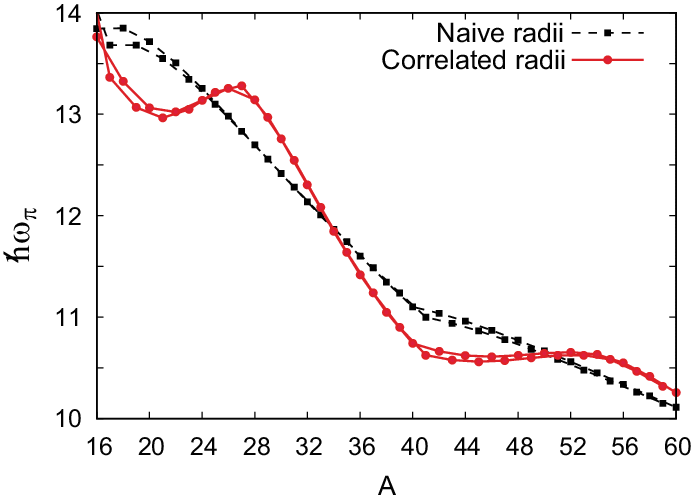}
\end{center}
\caption{\label{fig:hbw}(color online) Values of $\hbw_{\pi}$ (MeV)
  for $T=0$ and $1/2$ extracted from Eq.~(\ref{hbw}) using
  $\rho_{\pi}$ from Eqs.~(\ref{radii},\ref{lambda}) with
  parameters~$g=0.985,\, \rho_0=0.944,\,
  \lambda=5.562,\,\upsilon=0.368,\, \zeta=0.8$, for the
  correlated radii and $\lambda=0.$ for naive radii. All units in fm
  except $g$.}
\end{figure}
Two variants are chosen: $\lambda=0$ in Eqs.~(\ref{radii}) (the naive
fit) and $\lambda\ne 0$---the correlated fit---leading to the
interesting pattern in Fig~\ref{fig:hbw}. Its meaning may not be
evident at first, but clarification comes in
Fig.~\ref{fig:CaK}---showing the isotope shifts of the K and Ca
isotopes, including recent measures~\cite{Kradii,Caradii}---which make
it clear that Duflo's $\lambda$ term has a deep physical grounding:
The abrupt raise of radii after $A=47$ \ie the $N=28$ is an open
problem~\cite{Kradii,Caradii}, so far only qualitatively explained by
relativistic mean field calculations~\cite{RMF}. Fig.~\ref{fig:CaK}
suggests a very simple solution: the raise is due to the filling of
{\em huge} $p_{3/2}$ orbits. As the filling occurs for neutron orbits,
and the shift measures the behavior of proton orbits, isovector
polarizability must be at work here: if one fluid increases in size,
the other fluid must follow suit. The operation of the $\lambda$ term
does not depend on $\zeta$, which may take any value, but must be
fairly constant. To learn some more about the nature of $s_{1/2}$ and
$p_{3/2}$ which seem (are) responsible for the elegant undulating
patterns in Fig.~\ref{fig:hbw}, we examine the single particle and
single hole states built on \A{16}O and \A{40}Ca.

\begin{table}[t]
\caption{\label{tab:radii}MDE and MED $\Delta E$ for $T=1/2$ mirror
  nuclei of mass $A$, $\hbw_{\pi,\nu}$ in MeV and the corresponding
  skin parameters and radii in fm. Note that the radii correspond to
  the $N>Z$ nuclei, they are interchanged for the mirror
  partners. Experimental and calculated $\Delta E$ values coincide by
  construction. Interaction N3LO~\cite{n3lo} with cutoff $\Lambda=2$fm$^{-1}$.}
\begin{tabular*}{\linewidth}{@{\extracolsep{\fill}}ccccccccc}
\hline
\hline
$A$& $J^\pi$& $\Delta E$ &$\hbw_\nu$& $\hbw_\pi$& $\zeta$& $\Delta r_{\nu\pi}$& $r_\pi$& $r_\nu$ \\ 
\hline
15& $1/2^-$& 3.537& 14.55& 14.62& 0.358& 0.025& 2.507& 2.532 \\ 
  & $3/2^-$& 3.389& 14.39& 14.66& 0.609& 0.043& 2.503& 2.547 \\
\hline
17& $5/2^+$& 3.543& 13.62& 13.38& 0.906& 0.056& 2.641& 2.697 \\
  & $1/2^+$& 3.167& 12.86& 13.51& 2.367& 0.147& 2.628& 2.776 \\
\hline
39& $3/2^+$& 7.307& 10.97& 10.91& 0.258& 0.007& 3.361& 3.368 \\
  & $1/2^+$& 7.253& 10.90& 10.89& 0.523& 0.014& 3.365& 3.379 \\
\hline
41& $7/2^-$& 7.278& 10.78& 10.63& 0.610& 0.015& 3.422& 3.437 \\
  & $3/2^-$& 7.052& 10.61& 10.59& 1.513& 0.038& 3.427& 3.465 \\
  & $1/2^-$& 7.129& 10.61& 10.59& 1.482& 0.037& 3.428& 3.465 \\
  & $5/2^-$& 7.351& 10.75& 10.61& 0.702& 0.018& 3.424& 3.442 \\
  & $5/2^-$& 7.338& 10.75& 10.61& 0.725& 0.018& 3.427& 3.442 \\
\hline
\hline
\end{tabular*}
\end{table} 

\begin{figure}[ht]
\begin{center}
 \includegraphics[width=0.48\textwidth]{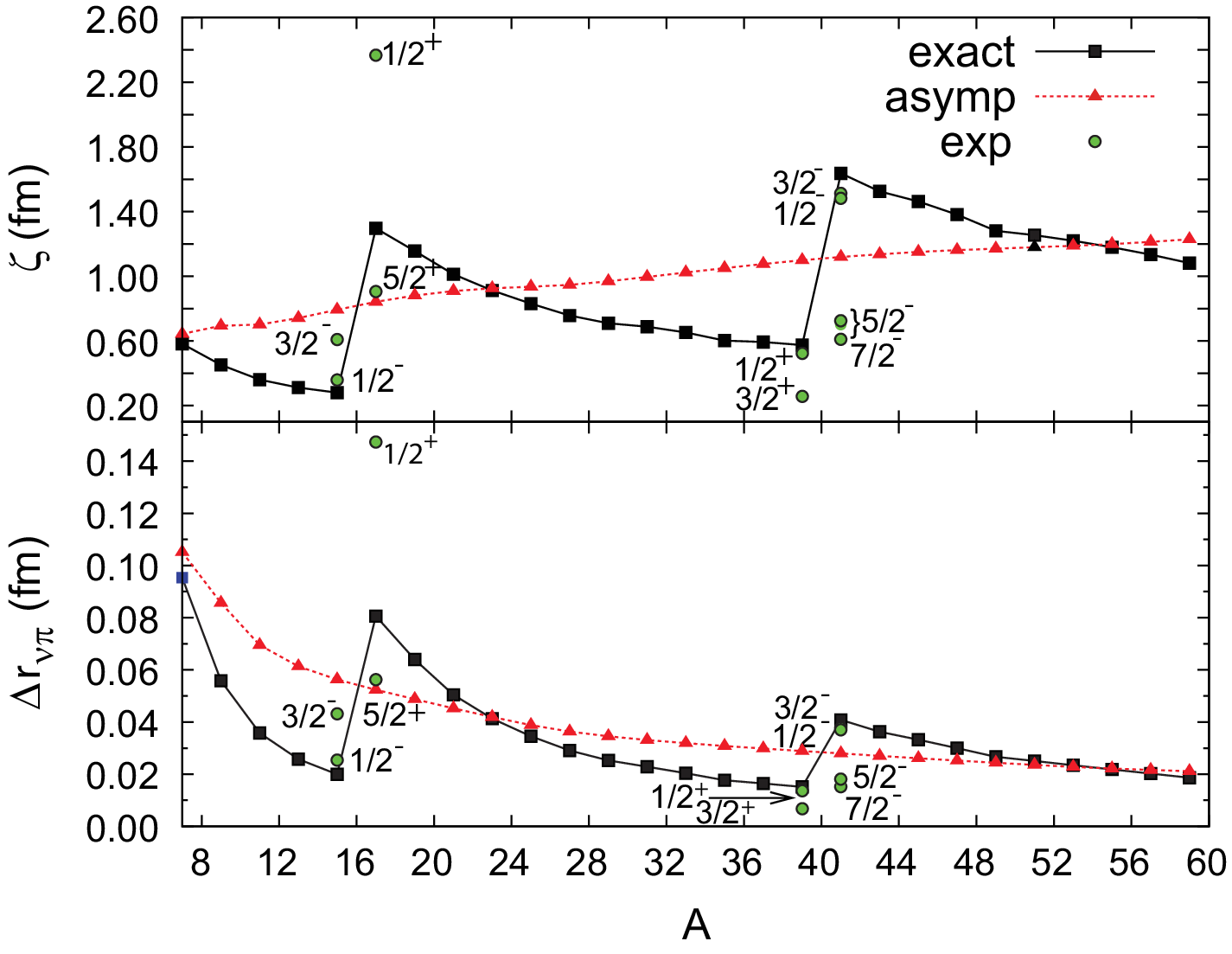}
\end{center}
\caption{\label{fig:zeta} (color online) Values of $\zeta$ and $\Delta
  r_{\nu\pi}$ from Table~\ref{tab:radii} (dots labeled exp) compared
  with those obtained for $\hbw_{\nu}=\hbw_{\pi}$ form
  Eqs.~(\ref{hbw}) (labeled exact) and (\ref{hbwasymp}) (labeled
  asymp).}
\end{figure}

 Results are given in Table~\ref{tab:radii} and
 Fig.~\ref{fig:zeta}. The values of $\zeta$ have been adjusted so as
 to obtain the observed energies. In the figure, the calculated
 $\zeta$ and $\Delta r_{\nu\pi}$ are compared with those obtained
 under the $\hbw_{\nu}=\hbw_{\pi}$ (naive shell model) assumption,
 expected to produce too large skins. However, because of the
 pronounced shell effects exhibited in the plots, for the hole states
 \ie $A=15$ and 39 the skins remain moderate or small. A few comments:

$A=15$. Independently of the $\zeta$ values, $\hbw_{\nu}<\hbw_{\pi}$
rules out an isovector polarization mechanism. As there is no simple
argument to treat these orbits as ``halo'', we prefer to leave the
question open.

$A=17$. A reasonable value of $\zeta$ solves the NSA for $d_{5/2}$. The
$s_{1/2}$ orbit is truly large: its rms radius is about 1.2 fm larger
than its  $d_{5/2}$ counterpart. No doubt about its halo nature.  

$A=39$. Here we find that $s_{1/2}$ is no longer gigantic, but large
enough to keep some memory of its halo status.

$A=41$. Most interesting. NSA is solved for $f_{7/2}$ via a reasonable
$\zeta$ very close to what is demanded by the lowest observed pair of
$f_{5/2}$ candidates which have only a fraction of the spectroscopic
strength. Both $p_{3/2}$ and $p_{1/2}$ are accommodated by the same
$\zeta$ and have pronounced halo nature. Their rms radii exceed those
of the $f$ orbits by some 0.7 fm. Interestingly, orbits of the same
$l$ have the same behavior.

\vspace{4pt}  

Old problems come back under new guises: the NSA as neutron skins, the
TES as halo orbits associated to subtle shell effects detected in
isotope shifts.

Neutron skins are difficult to measure experimentally. Recent progress
has been made~\cite{Pbskin} and their connection with the isovector
dipole polarizability $\alpha_D$ have led to reliable
estimates~\cite{rocam}. Theoretically the problem is much simpler. It
is subsumed by isovector monopole polarizability~\cite{czech}, or for
Skyrme type functionals by control of the symmetry
energy~\cite{gomezm,brownEOS}. As noted after Table~\ref{tab:estim},
several calculations appear to reproduce skins well.

Halo orbits are another matter: no existing
calculation~\cite{Kradii,Caradii} explains the observed isotope shifts
as done in Fig.~\ref{fig:CaK}. We have interpreted the result as due
to an increase in size of a $p$ orbit. We have also learned from
Table~\ref{tab:radii} and Fig.~\ref{fig:zeta} that $s_{1/2}$ and
$p_{3/2}$ are so huge that they could be viewed as halo orbits in
$A=17$ and 41, but their influence extends well beyond. We have also
learned that at $A=39$, $s_{1/2}$ is no longer huge. We expect to
learn much about its evolution through full MED and MDE configuration
mixing calculations now under way.

We close by proposing an alternative to the use of Eq.~(\ref{lambda})
to represent shell effects: 
\begin{gather}\label{r2pi}
\langle r_{\pi}^2 \rangle=\frac{41.47}{\hbw_{\pi}}\sum_i
z_i(p_i+3/2+\delta_i)/Z
\end{gather}   
where $\hbw_{\pi}$ is now the ``naive'' estimate using
Eq.~(\ref{radii}) alone and the $\delta_i$ corrections to the
oscillator values replace the $\lambda$ term. Eq.~(\ref{r2pi}) could
be useful in interpreting the stucture of isotope shifs as reflecting
orbital occupancies associated to given orbital radii.

\end{document}